\documentstyle[12pt,epsf]{article}
\title{Dynamics of transverse correlations
       \protect\\
       in the spin-$\frac{1}{2}$ isotropic $XY$ chain
       \protect\\
       with a correlated Lorentzian disorder}
\author{Oleg Derzhko$^{\dagger,\ddagger}$
        and Taras Krokhmalskii$^{\dagger}$\\
\small   {\em {$^{\dagger}$Institute for Condensed Matter Physics,}}\\
\small   {\em {1 Svientsitskii St., L'viv-11, 290011, Ukraine}}\\
\small   {\em {$^{\ddagger}$Chair of Theoretical Physics,
               Ivan Franko L'viv State University,}}\\
\small   {\em {12 Drahomanov St., L'viv-5, 290005, Ukraine}}}
\date{\today}

\begin{document}

\maketitle

\begin{abstract}
Using a numerical approach we examine
the dynamics of $zz$ correlations
for the spin-$1\over2$ isotropic $XY$ chain
with random Lorentzian intersite coupling
and transverse field at sites
that depends linearly on the neighbouring couplings.
We study in detail
the wave vector- and frequency-dependent $zz$ structure factor
at different values of Hamiltonian parameters and temperature.
We discuss the changes in the frequency profiles
of the $zz$ dynamic structure factor
caused by the introduced correlated disorder.
\end{abstract}

\vspace{1cm}


\vspace{1cm}


\vspace{1mm}

\noindent
{\bf Postal address:}\\
{\em
Dr. Oleg Derzhko (corresponding author)\\
Dr. Taras Krokhmalskii\\
Institute for Condensed Matter Physics\\
1 Svientsitskii St., L'viv-11, 290011, Ukraine\\
Tel: (0322) 42 74 39

\noindent
\hspace{8mm}
(0322) 76 09 08\\
Fax: (0322) 76 19 78\\
E-mail: derzhko@icmp.lviv.ua

\noindent
\hspace{14mm}
krokhm@icmp.lviv.ua}

\clearpage

\renewcommand\baselinestretch {1.65}
\large\normalsize

Quantum spin chains with randomness have attracted a great deal of interest
over the last decades. Apparently the simplest example of such a system
is a random spin-$\frac{1}{2}$ $XY$ chain
the study of which is essentially simplified due to the fact
that with the help of the Jordan-Wigner transformation
it can be presented in terms of spinless noninteracting fermions.
Starting from the early 70s several types of random
spin-$\frac{1}{2}$ $XY$ chains were discussed by using fermionization
and Dyson's and Lloyd's models of disorder, as well as a numerical
approach [1-7].
The interest in random
spin-$\frac{1}{2}$ $XY$ chains has been strongly renewed in the 90s
in view of the study of generic features of quantum phase transitions
in disordered systems.
As an example we mention here an exhaustive study of 
the transverse Ising chain
by both the renormalization group and numerical means [8-10],
and exact analytical and numerical treatment of
$XY$ chain [11-13].

In this paper, we deal with the
spin-$\frac{1}{2}$ isotropic $XY$ chain with random
Lorentzian exchange coupling and transverse field that depends linearly
on the surrounding exchange couplings.
Such a model has recently been examined
in some detail [14].
The method developed by John and Schreiber [15]
allowed one to calculate exactly
the random-averaged density of magnon states for that model and, 
as a result,
to study rigorously its thermodynamics.
The most interesting property that the model
with a correlated disorder exhibits is the appearance of
a nonzero averaged transverse magnetization
at the zero averaged transverse field [14,16,17].
This is conditioned by the change in the density of states
due to the correlated
disorder. Namely, the numbers of
``magnons" with negative and positive energies
at the zero averaged transverse field become not equal to each other.
Unfortunately,
the obtained analytical results pertain only to thermodynamics.
The aim of the present paper is to
study the effects of a correlated disorder on
dynamics of transverse spin correlations,
examining for this purpose
the $zz$ dynamic
structure factor.
To reveal
the effects of the correlated off-diagonal and diagonal disorder
it is necessary to analyse also the model with
independent
random Lorentzian
exchange couplings and transverse fields.
Such a study of dynamic properties requires
the calculation of the $zz$ time-dependent spin
correlation functions and concerns the dynamics of the spin model
conditioned by the exciting of only two magnons.
Apparently, the evaluation
even of the
simplest $zz$ time-dependent spin correlation functions
cannot be performed analytically,
however, it can be done numerically,
making use of the developed earlier finite-chain calculation scheme [18]
(similar approaches are described in Refs. [9,10,19,20]).

It should be stressed that models with the correlated disorder naturally
arise while describing materials with the topological disorder.
On the other hand,
dynamic measurements are the basic experimental
techniques in the study of such compounds.
Although
there are a few examples of real materials which are well
described by a one-dimensional spin-$\frac{1}{2}$ isotropic $XY$ model
(for instance, PrCl$_3$ [21])
the presented
below exact
numerical results do not pertain to any particular compound.
Nevertheless, they
still may be
of much use for understanding
the possible changes in observable
quantities caused by the correlated disorder
and can help to link the theoretical predictions and experimental data.

Following Ref. 14
we consider a linear
isotropic $XY$ chain of $N$ spins $\frac{1}{2}$
in a transverse field governed by the Hamiltonian
\begin{eqnarray}
H=\sum_{n=1}^N\Omega_ns_n^z
+\sum_{n=1}^{N-1}J_n\left(s_n^xs_{n+1}^x+s_n^ys_{n+1}^y\right)
\nonumber\\
=\sum_{n=1}^N\Omega_n\left(s_n^+s_n^--\frac{1}{2}\right)
+\sum_{n=1}^{N-1}\frac{J_n}{2}
\left(s_n^+s_{n+1}^-+s_n^-s_{n+1}^+\right)
\end{eqnarray}
where $\Omega_n$ is the transverse field at site $n$
and $J_n$ is the exchange coupling between the sites $n$ and $n+1$.
The $J_n$ are taken to be independent random variables
with the Lorentzian probability distribution
\begin{eqnarray}
p(J_n)=\frac{1}{\pi}
\frac{\Gamma}{\left(J_n-J_0\right)^2+\Gamma^2}
\end{eqnarray}
where $J_0$ is the mean value of the exchange coupling and $\Gamma$
is the width of distribution that controls the strength of the disorder.
For the correlated off-diagonal and diagonal disorder the transverse fields
are related to the intersite couplings
and further it is assumed that
\begin{eqnarray}
\Omega_n-\Omega_0
=a\left(\frac{J_{n-1}-J_0}{2}+\frac{J_n-J_0}{2}\right)
\end{eqnarray}
where $\Omega_0$ is the averaged transverse field at site [22].
One can easily find the probability distribution for the random variable
$\Omega_n$
\begin{eqnarray}
p(\Omega_n)
=\int\limits_{-\infty}^{\infty}
\int\limits_{-\infty}^{\infty}
{\mbox{d}}J_{n-1}{\mbox{d}}J_{n}
p(J_{n-1})p(J_n)\delta
\left(
\Omega_n-\Omega_0-
a\left(\frac{J_{n-1}-J_0}{2}+\frac{J_n-J_0}{2}\right)
\right)
\nonumber\\
=\frac{1}{\pi}
\frac{\vert a\vert\Gamma}
{(\Omega_n-\Omega_0)^2+(\vert a\vert\Gamma)^2},
\end{eqnarray}
i.e. $\Omega_n$ appears to be a Lorentzian random variable
with the mean value $\Omega_0$
and the width of the distribution $\vert a\vert\Gamma$.

Mapping spin model (1) with the help of the
Jordan-Wigner transformation
onto spinless noninteracting fermions one obtains the Hamiltonian
$$H=-\frac{1}{2}\sum_{n=1}^N\Omega_n
+\sum_{n,m=1}^Nc_n^+A_{nm}c_m$$
with
$$A_{nm}=\Omega_n\delta_{nm}
+\frac{1}{2}J_n\delta_{m,n+1}
+\frac{1}{2}J_{n-1}\delta_{m,n-1}$$
that can be diagonalized by the transformation
$\eta_\kappa^+=\sum_{j=1}^Ng_{kj}c^+_j,$
$\eta_\kappa=\sum_{j=1}^Ng_{kj}c_j,$
where
\begin{eqnarray}
\sum_{j=1}^Ng_{kj}A_{jm}=\Lambda_kg_{km},
\;\;\;
\sum_{j=1}^Ng_{kj}g_{qj}=\delta_{kq},
\;\;\;
\sum_{k=1}^Ng_{kj}g_{km}=\delta_{jm},
\end{eqnarray}
with the result
$H=\sum_{\kappa=1}^N\Lambda_{\kappa}\left(\eta_{\kappa}^+\eta_{\kappa}
-\frac{1}{2}\right)$.
Since
$s_j^z=-\frac{1}{2}\varphi^+_j\varphi^-_j,$
where
$\varphi^\pm_j
=\sum_{p=1}^Ng_{pj}
\left(\eta^+_p\pm \eta_p\right),$
the calculation of
the $zz$ time-dependent spin correlation functions
reduces to exploiting
the Wick-Bloch-de Dominicis theorem
with the outcome
\begin{eqnarray}
\langle s_j^z(t)s_{j+n}^z\rangle
-\langle s_j^z\rangle\langle s_{j+n}^z\rangle=
\frac{1}{4}
\left(
-\langle\varphi_j^+(t)\varphi_{j+n}^+\rangle
\langle\varphi_{j}^-(t)\varphi_{j+n}^-\rangle
+\langle\varphi_j^+(t)\varphi_{j+n}^-\rangle
\langle\varphi_{j}^-(t)\varphi_{j+n}^+\rangle
\right)
\end{eqnarray}
where the elementary contractions read
\begin{eqnarray}
\langle\varphi_j^+(t)\varphi_m^+\rangle=
-\langle\varphi_j^-(t)\varphi_m^-\rangle=
\sum_{p=1}^N
g_{pj}g_{pm}
\left(
\frac{{\mbox{e}}^{{\mbox{i}}\Lambda_pt}}{1+{\mbox{e}}^{\beta\Lambda_p}}
+
\frac{{\mbox{e}}^{-{\mbox{i}}\Lambda_pt}}{1+{\mbox{e}}^{-\beta\Lambda_p}}
\right),
\nonumber\\
\langle\varphi_j^+(t)\varphi_m^-\rangle=
-\langle\varphi_j^-(t)\varphi_m^+\rangle=
-\sum_{p=1}^N
g_{pj}g_{pm}
\left(
\frac{{\mbox{e}}^{{\mbox{i}}\Lambda_pt}}{1+{\mbox{e}}^{\beta\Lambda_p}}
-
\frac{{\mbox{e}}^{-{\mbox{i}}\Lambda_pt}}{1+{\mbox{e}}^{-\beta\Lambda_p}}
\right).
\end{eqnarray}
Formulae (5) - (7) are the basic ones for the
the presented below
numerical study of the dynamic properties
of spin model (1) - (4).
For the given realization of random couplings (2)
and corresponding transverse fields (3)
or for the given realizations of
random couplings (2)
and random transverse fields (4)
one must first calculate
the eigenvalues and
eigenvectors of matrix $\bf A$ (5).
Knowing its eigenvalues and eigenvectors
one immediately obtains
elementary contractions (7) and,
therefore, the $zz$ time-dependent
spin correlation functions  (6)
that are directly related to the
$zz$ dynamic structure factor or susceptibility.
Usually one is interested in the random-averaged quantities that
come out as the result of averaging the computed quantities
over many random realizations.
The random-averaged quantities will be overlined.
More details on the finite-chain calculation scheme
can be found in Ref. [18].

In what follows we shall discuss
the dynamics of transverse spin correlations in spin
model (1) - (4), calculating for this purpose
the transverse dynamic structure factor
\begin{eqnarray}
\overline{S_{zz}(\kappa,\omega)}
=\sum_{n=1}^N
{\mbox{e}}^{{\mbox{i}}\kappa n}
\int\limits_{-\infty}^{\infty}{\mbox{d}}t
{\mbox{e}}^{-\epsilon\mid t\mid}
{\mbox{e}}^{{\mbox{i}}\omega t}
\left[
\overline{\langle s_j^z(t)s_{j+n}^z\rangle}
-
\overline{\langle s_j^z\rangle\langle s_{j+n}^z\rangle}
\right]
\nonumber\\
=
2
\sum_{n=0,\pm 1,\pm 2, \ldots}
{\mbox{e}}^{{\mbox{i}}\kappa n}
{\mbox{Re}}
\int\limits_0^{\infty}{\mbox{d}}t
{\mbox{e}}^{{\mbox{i}}(\omega+{\mbox{i}}\epsilon)t}
\left[
\overline{\langle s_j^z(t)s_{j+n}^z\rangle}
-
\overline{\langle s_j^z\rangle\langle s_{j+n}^z\rangle}
\right],
\;\;\;\epsilon\rightarrow +0.
\end{eqnarray}
The transverse dynamic structure factor (8)
for a certain random realization
with the help of Eqs. (6), (7) can be rewritten
in the following form
\begin{eqnarray}
S_{zz}(\kappa,\omega)
=2\pi
\sum_{n=0,\pm 1,\pm 2,\ldots}
{\mbox{e}}^{{\mbox{i}}\kappa n}
\sum_{p,q=1}^N
g_{pj}g_{p,j+n}g_{qj}g_{q,j+n}
\frac{\delta(\omega+\Lambda_p-\Lambda_q)}
{\left(1+{\mbox{e}}^{\beta\Lambda_p}\right)
\left(1+{\mbox{e}}^{-\beta\Lambda_q}\right)}
\end{eqnarray}
[23].
For a uniform cyclic infinite chain
($\Omega_n=\Omega_0$, $J_n=J_0$)
the corresponding result reads
\begin{eqnarray}
S_{zz}(\kappa,\omega)
=\int\limits_{-\pi}^{\pi}
{\mbox{d}}\kappa^{\prime}
\frac{\delta(\omega
+\Lambda_{\kappa^{\prime}}-\Lambda_{\kappa^{\prime}-\kappa})}
{\left(1+{\mbox{e}}^{\beta\Lambda_{\kappa^{\prime}}}\right)
\left(1+{\mbox{e}}^{-\beta\Lambda_{\kappa^{\prime}-\kappa}}\right)},
\nonumber\\
\Lambda_{\kappa}=\Omega_0+J_0\cos\kappa.
\end{eqnarray}
Evaluating the integral in Eq. (10) one gets
the following expression for
the transverse
dynamic structure factor
of the uniform cyclic infinite chain
\begin{eqnarray}
S_{zz}(\kappa,\omega)
=
\left\{
\begin{array}{ll}
\sum_{\kappa^{\star}}
\frac{1}{\vert -J\sin\kappa^{\star}+J\sin(\kappa^{\star}-\kappa)\vert}
\frac{1}{\left(1+{\mbox{e}}^{\beta\Lambda_{\kappa^{\star}}}\right)
\left(1+{\mbox{e}}^{-\beta\Lambda_{\kappa^{\star}-\kappa}}\right)},
& {\mbox{if}}\;\;\;\omega\le2\vert J\sin\frac{\kappa}{2}\vert,
\\
0,
&
{\mbox{otherwise}}
\end{array}
\right.
\end{eqnarray}
with
\begin{eqnarray}
\kappa^{\star}=\left\{
\frac{\kappa}{2}+\arcsin\frac{\omega}{2J\sin\frac{\kappa}{2}},\;\;\;
\frac{\kappa}{2}+\pi-\arcsin\frac{\omega}{2J\sin\frac{\kappa}{2}}
\right\}.
\end{eqnarray}

Let us describe the sketched above
numerical analysis of the dynamic properties
of the considered spin system in more detail.
To understand the accuracy of the numerical results we performed
many additional calculations.
In Figs. 1a, 1b
one can see the time dependence of the autocorrelation function
$\langle s_{\frac{N}{2}}^z(t)s_{\frac{N}{2}}^z\rangle
-\langle s_{\frac{N}{2}}^z\rangle^2$
for $N=150$ and $N=300$, respectively.
The plots demonstrate the finite size effects
that appear at $t\approx 150$ for $N=150$ and
$t\approx 300$ for $N=300$.
The same effects can be seen in the time dependence of
$\langle s_{150}^z(t)s_{250}^z\rangle
-\langle s_{150}^z\rangle\langle s_{250}^z\rangle$
computed for $N=300$, which is depicted in Fig. 1c.
This correlation function appears with a delay
at $t\approx 100$
and at times $t\approx 200$ exhibits a time behaviour
influenced by the finite size of the chain considered.
Figs. 2a and 2b demonstrate the difference between the sums
$$
\sum_{n=0,\pm1,\ldots,\pm n^{\star}}
{\mbox{e}}^{{\mbox{i}}\kappa n}
\left[
\langle s_{150}^z(t)s_{150+n}^z\rangle
-\langle s_{150}^z\rangle\langle s_{150+n}^z\rangle
\right]$$
($N=300$)
with different $n^{\star}$.
Because of increasing the time of delay in 
the appearance of correlation
functions with large $n$, it is necessary to take a sufficiently
large number of terms in the sum to reproduce correctly its time
behaviour at large times
(compare Figs. 2a and 2b that correspond to
$n^{\star}=50$ and $n^{\star}=100$, respectively).
In practice
one must restrict
the calculations
of the time-dependent spin correlation functions to
some finite time of cut-off $t_c$
that generates wiggles in the frequency dependence
of the dynamic structure factor.
The wiggles can be removed by increasing the time of the cut-off
to such values at which
the spin correlations are already small enough
or by increasing the value of $\epsilon$ that
smoothes the frequency profiles.
Fig. 3 demonstrates how the numerical results approach the exact ones
with increasing $t_c$ and decreasing $\epsilon$.
Finally, let us add that
acting in the described manner we reproduce the analytical results
for the transverse dynamic susceptibility in a non-random case
[24] (see also [25-27]).
Figs. 4a - 4c
show that in the random case one may take
much smaller
values of $t_c$, since the random-averaged sum of correlation functions
that yields $\overline{S_{zz}(\kappa,\omega)}$
decays essentially faster than in the non-random case
(compare Figs. 4c and 2).
Besides, with increasing 
the number of random realizations the
resulting random-averaged sum of correlation functions
becomes more regular (compare Figs. 4a, 4b and 4c).
In our numerical calculations we considered
spin-$\frac{1}{2}$ transverse isotropic $XY$ chains
of $N=200, 300, 1800$ spins with $J_0=-1$
(the results for $\overline{S_{zz}(\kappa,\omega)}$
do not depend on the sign of exchange coupling),
$\Omega_0=0, 0.25, 0.5, 0.75, 0.99$ and $\Gamma=0,\;0.1$
at low ($\beta=1000$)
and high ($\beta=1$) temperatures.
We computed correlation functions
$\langle s_{\frac{N}{2}}^z(t)s_{\frac{N}{2}+n}^z\rangle
-\langle s_{\frac{N}{2}}^z\rangle\langle s_{\frac{N}{2}+n}^z\rangle$
with $n$ up to $n^{\star}=80, \ldots, 800$
for the times up to $t_c=150, \ldots, 1200$, put $\epsilon=0.005$
and averaged the $zz$ dynamic structure factor
at least over 21000 random
realizations.
In the non-random case we used exact formulae (11), (12).
The obtained within the frames of the described scheme results
for the transverse dynamic structure factor of the non-random chain
and the chains with a correlated
and non-correlated
Lorentzian disorder are presented
in Figs. 5 and 6, respectively.

Let us turn to the discussion of the obtained results.
At first 
let us consider a non-random case (formulae (10) - (12), Fig. 5).
To explain the observed frequency profiles one must take into account
the fact
that they reflect the dynamic properties of the magnetic chain
conditioned by the exciting of two magnons
with energies
$\Lambda_{{\kappa}^{\prime}}=\Omega_0+J_0\cos\kappa^{\prime}$
and
$\Lambda_{{\kappa}^{\prime\prime}}=\Omega_0+J_0\cos\kappa^{\prime\prime}$
for which
$\omega=-\Lambda_{{\kappa}^{\prime}}
+\Lambda_{{\kappa}^{\prime\prime}}$
and
$\kappa^{\prime\prime}={\kappa}^{\prime}-{\kappa}$.
Besides, at $T=0$
($\beta\to\infty$)
the magnon energies
must have the corresponding signs
due to the Fermi factors involved into (10),
namely
$\Lambda_{{\kappa}^{\prime}}<0$,
$\Lambda_{{\kappa}^{\prime\prime}}>0$.
Consider, for example, $S_{zz}(\frac{\pi}{4},\omega)$
for $\Omega_0=0.5$
(the curve for $\kappa=\frac{\pi}{4}$ in Fig. 5c
and the dashed curve in Fig. 6b).
Since $\Lambda_{\kappa}=0.5-\cos\kappa$
for $T=0$,
because of the Fermi factors,
$\kappa^{\prime}$ may vary only from
$-\frac{\pi}{3}$ to $-\frac{\pi}{12}$
and, hence,
$\kappa^{\prime}-\frac{\pi}{4}$ varies from
$-\frac{7\pi}{12}$ to $-\frac{\pi}{3}$,
whereas
for non-zero temperature
$\kappa^{\prime}$ varies from $-\pi$ to $\pi$.
The value of
$-\Lambda_{\kappa^{\prime}}+\Lambda_{\kappa^{\prime}-\frac{\pi}{4}}
=-2\sin{\frac{\pi}{8}}\sin{\left(\kappa^{\prime}-\frac{\pi}{8}\right)}$
is shown in Fig. 7a.
Evidently, at $T=0$,
the lower frequency at which the non-zero value of
$S_{zz}(\frac{\pi}{4},\omega)$ appears,
$\omega_l$, is equal to
$\cos{\frac{\pi}{12}}-\cos{\frac{\pi}{3}}=0.46592\ldots$
and the upper frequency
$\omega_u$,
after which
$S_{zz}(\frac{\pi}{4},\omega)$ disappears,
is equal to
$\cos{\frac{\pi}{3}}-\cos{\frac{7\pi}{12}}=0.75881\ldots\;$;
for non-zero temperature
$\omega_l=0$
and
$\omega_u=2\sin\frac{\pi}{8}=0.76536\ldots$
(see Fig. 7a).
The value of
$S_{zz}(\frac{\pi}{4},\omega)$
is determined by the value of the slope of the curve depicted in Fig. 7a
and this explains why
$S_{zz}(\frac{\pi}{4},\omega_l)<S_{zz}(\frac{\pi}{4},\omega_u)$
at $T=0$,
as well as why
$S_{zz}(\frac{\pi}{4},\omega_u)\to\infty$
at non-zero temperature.
Note that, as it can be seen from Fig. 7a, the value of
$S_{zz}(\frac{\pi}{4},\omega)$ is determined by two pairs of magnons
that satisfy the conditions
$\omega=-\Lambda_{{\kappa}^{\prime}}
+\Lambda_{{\kappa}^{\prime\prime}}$
and
$\kappa^{\prime\prime}={\kappa}^{\prime}-{\kappa}$,
however, at $T=0$, because of the additional requirement
$\Lambda_{{\kappa}^{\prime}}<0$,
$\Lambda_{{\kappa}^{\prime\prime}}>0$,
only one pair of magnons can contribute to
$S_{zz}(\frac{\pi}{4},\omega)$.

Consider further, for example,
$S_{zz}(\frac{2\pi}{3},\omega)$
for $\Omega_0=0.5$
(the curve for $\kappa=\frac{2\pi}{3}$ in Fig. 5c
and the dashed curve in Fig. 6d).
For $T=0$, $\kappa^{\prime}$ may vary only from
$-\frac{\pi}{3}$
to
$\frac{\pi}{3}$,
whereas for non-zero temperature
$\kappa^{\prime}$ varies from
$-\pi$
to
$\pi$.
The value of
$-\Lambda_{\kappa^{\prime}}+\Lambda_{\kappa^{\prime}-\frac{2\pi}{3}}
=-\sqrt{3}\sin\left(\kappa^{\prime}-\frac{\pi}{3}\right)$
is shown in Fig. 7b.
As it can be seen from this figure, for $T=0$,
as well as for non-zero temperature $\omega_l=0$
and $\omega_u=\sqrt{3}=1.73205\ldots\;$.
Besides,
$S_{zz}(\frac{2\pi}{3},\omega_u)\to\infty$
for any temperature.
The abrupt change in
$S_{zz}(\frac{2\pi}{3},\omega)$
at $T=0$ at $\omega=1.5$ is due to the fact that
for lower frequencies only one pair of magnons, 
because of the Fermi factors,
contributes to
$S_{zz}(\frac{2\pi}{3},\omega)$,
whereas for higher frequencies two pairs of magnons are involved in forming
$S_{zz}(\frac{2\pi}{3},\omega)$.
At non-zero temperature
$S_{zz}(\frac{2\pi}{3},\omega)$
is always conditioned by two pairs of magnons.

Let us pass to random models
(formula (9), Fig. 6).
For such models the transverse dynamic structure factor is again
conditioned by two magnons
$\Lambda_p$ and $\Lambda_q$,
for which
$\omega=-\Lambda_p+\Lambda_q$
and the quantity
$$
\Delta(\kappa,p,q)
=
\sum_{n=0,\pm 1,\pm 2,\ldots}
{\mbox{e}}^{{\mbox{i}}\kappa n}
g_{pj}g_{p,j+n}g_{qj}g_{q,j+n}
$$
has a non-zero value.
Besides, at $T=0$,
$\Lambda_p<0$,
$\Lambda_q>0$.
Apparently, 
there is no simple rigorous explanation for the behaviour
of the random-averaged frequency profiles depicted in Fig. 6.
However, it is easy to note that the kind of naive reasoning
presented below does work for such a case.
Consider at first the case
$\kappa=\frac{\pi}{4}$ at low temperature.
As it was shown above,
the non-zero value of
$S_{zz}(\frac{\pi}{4},\omega_l)$
in a non-random case was conditioned by two magnons with the energies
$0.5-\cos\frac{\pi}{12}=-0.46592\ldots$
and
$0.5-\cos\frac{\pi}{3}=0$, respectively.
As it can be seen in Fig. 8,
where the random-averaged
densities of magnon states
$\overline{\rho(E)}
=\overline{\frac{1}{N}\sum_{k=1}^N\delta(E-\Lambda_k)}$
obtained from (5)
are depicted,
such a pair of ``magnons" does exsist for
$a=-1.01$ (Fig. 8a)
and does not exsist for
$a=1.01$ (Fig. 8b)
or for the case of a non-correlated disorder (Fig. 8c).
This observation is in agreement with the changes in the frequency
profile due to different types of disorder
shown in Fig. 6b
(compare the curves 1, 4
and 2, 3 at $\omega_l$).
In the non-random case the non-zero value of
$S_{zz}(\frac{\pi}{4},\omega_u)$
was conditioned by two magnons with the energies
$0.5-\cos\frac{\pi}{3}=0$
and
$0.5-\cos\frac{7\pi}{12}=0.75881\ldots\;$, respectively.
From Fig. 8 one can see that the density of states for such magnons
is diminished because of the disorder that agrees with the changes in
$S_{zz}(\frac{\pi}{4},\omega_u)$
shown in Fig. 6b.

Consider further the case
$\kappa=\frac{2\pi}{3}$ at low temperature.
For the non-random case
$S_{zz}(\frac{2\pi}{3},\omega_l)$
arises due to two magnons
$\Lambda_{\frac{\pi}{3}}$ and $\Lambda_{-\frac{\pi}{3}}$
with the zero energy.
As it can be seen in Fig. 8,
the disorder affects
the density of states
at $E=0$ for $a=1.01$ and the non-correlated disorder more than for
$a=-1.01$
that agrees with the changes in the frequency profile
(Fig. 6d).
$S_{zz}(\frac{2\pi}{3},\omega_u)$
is formed by two magnons with the energies
$0.5-\cos\frac{\pi}{6}=-0.36602\ldots$ and
$0.5-\cos\frac{5\pi}{6}=1.36602\ldots\;$, respectively.
The density of magnon states for
such energies is more diminished for the non-correlated disorder
(Fig. 8c) than for the correlated one
(Figs. 8a, 8b)
which agrees with the smaller value of
$\overline{S_{zz}(\frac{2\pi}{3},\omega_u)}$
in the former case in comparison with the latter.

To summarize, we extended the consideration of the spin-$\frac{1}{2}$
isotropic $XY$ chain with a correlated Lo\-ren\-tzian disorder
presented in Ref. 14,
examining numerically the dynamics of transverse spin correlations.
We obtained the frequency dependences of
the transverse dynamic structure factor
at different values of the
wave vector and temperature.
We found the possible influences of the correlated disorder on the
frequency profiles of the transverse dynamic structure factor.
Within certain frequency regions
the introducing of the correlated disorder
may yield almost no changes in the value of
$\overline{S_{zz}(\kappa,\omega)}$,
whereas the non-correlated disorder always erodes the frequency profiles
of $\overline{S_{zz}(\kappa,\omega)}$.
The studied
possible influences of the correlated
disorder on the dynamic properties may be useful
for the analysis of experimental data
obtained in dynamic experiments
for quasi-one-dimensional spin-$\frac{1}{2}$
isotropic $XY$ compounds.

The authors are grateful to Prof. M. Shovgenyuk
for providing the possibility to perform the numerical calculations.
The paper was discussed at Magdeburg University
and Dortmund University.
O. D. is grateful to Prof. J. Richter and Prof. J. Stolze for their
warm hospitality.
He is also indebted to Mrs. Olga Syska for the financial support.

\clearpage

\noindent
{\bf List of figure captions}

\vspace{1cm}

\noindent
FIG. 1.
Time dependence of the transverse spin correlation functions
$r_{j,j+n}(t)={\mbox{Re}}
\left[\langle s_{j}^z(t)s_{j+n}^z\rangle
\right.
$
$
\left.
-\langle s_{j}^z\rangle\langle s_{j+n}^z\rangle\right]$
for $J_0=-1$, $\Omega_0=0.5$, $\Gamma=0$ at $\beta=1000$.
a) $N=150$, $j=j+n=75$;
b) $N=300$, $j=j+n=150$;
c) $N=300$, $j=150$, $j+n=250$.
In this figure and in Figs. 2, 4 we plotted only the real parts of
correlation functions since their imaginary parts exhibit
qualitatively the same behaviour.

\vspace{1cm}

\noindent
FIG. 2.
Time dependence of
$R_{\kappa}^{n^{\star}}(t)={\mbox{Re}}
\sum_{n=0,\pm1,\ldots,\pm n^{\star}}
{\mbox{e}}^{{\mbox{i}}\kappa n}
\left[
\langle s_{150}^z(t)s_{150+n}^z\rangle
-\langle s_{150}^z\rangle\langle s_{150+n}^z\rangle
\right],$
$\kappa=\pi$
for $J_0=-1$, $\Omega_0=0.5$, $\Gamma=0$, $N=300$ at $\beta=1000$
for $n^{\star}=50$ (a)
and $n^{\star}=100$ (b).

\vspace{1cm}

\noindent
FIG. 3.
$S_{zz}(\kappa,\omega)$
at $\kappa=\frac{\pi}{8}$ (1),
$\kappa=\frac{\pi}{4}$ (2),
$\kappa=\frac{2\pi}{3}$ (3),
$\kappa=\frac{3\pi}{4}$ (4),
$\kappa=\pi$ (5)
for the uniform chain
with $J_0=-1$, $\Omega_0=0.5$ at $\beta=1000$:
exact results (11), (12) (dashed curves)
versus numerical ones
(solid curves).
a)
$N=280$,
$j=140$,
$n^{\star}=100$,
$t_c=200$, $\epsilon=0.01$;
b)
$N=1800$,
$j=900$,
$n^{\star}=800$,
$t_c=1200$, $\epsilon=0.01$;
c)
$N=1800$,
$j=900$,
$n^{\star}=800$,
$t_c=1200$, $\epsilon=0.005$.

\vspace{1cm}

\noindent
FIG. 4.
Time dependence of
$\overline{R_{\kappa}^{n^{\star}}(t)},$
$\kappa=\pi$
for $J_0=-1$, $\Omega_0=0.5$,
$\Gamma=0.1$, $a=1.01$,
$N=300$, $n^{\star}=100$ at $\beta=1000$;
the random-averaged quantity comes as the result of
averaging over 1000 realizations (a),
11000 realizations (b),
and 21000 realizations (c).

\vspace{1cm}

\noindent
FIG. 5.
Frequency dependence of the transverse dynamic structure factor
(11), (12)
for
$J_0=-1$
and
different values of transverse field
$\Omega_0=0$ (a),
$\Omega_0=0.25$ (b),
$\Omega_0=0.5$ (c),
$\Omega_0=0.75$ (d),
$\Omega_0=0.99$ (e)
at different values of wave vector
$\kappa
=\frac{\pi}{8},
\frac{\pi}{4},
\frac{\pi}{2},
\frac{2\pi}{3},
\frac{3\pi}{4},
\pi$
(from left to right)
at $\beta=1000$.
At high temperatures the
frequency profiles of
$S_{zz}(\kappa,\omega)$
are the same for all values of $\Omega_0$ (f).

\vspace{1cm}

\noindent
FIG. 6.
Frequency dependence of the random-averaged transverse dynamic structure
factor (8) at different values of wave vector
$\kappa=\frac{\pi}{8}$ (a),
$\kappa=\frac{\pi}{4}$ (b),
$\kappa=\frac{\pi}{2}$ (c),
$\kappa=\frac{2\pi}{3}$ (d),
$\kappa=\frac{3\pi}{4}$ (e),
$\kappa=\pi$ (f)
for model (1) with
$J_0=-1,$
$\Omega_0=0.5,$
$\Gamma=0.1$
at $\beta=1000.$
1) correlated disorder (3) with $a=-1.01;$
2) correlated disorder (3) with $a=1.01;$
3) independent exchange couplings and transverse fields,
the latter are distributed according to
probability distribution (4) with
$\vert a\vert\Gamma=0.101;$
4) non-random case $\Gamma=0$ (dashed curves).

\vspace{1cm}

\noindent
FIG. 7.
$-\Lambda_{\kappa^{\prime}}
+\Lambda_{\kappa^{\prime}-\kappa}$
versus
$\kappa^{\prime}$
for the non-random chain with $J_0=-1$,
$\Omega_0=0.5$.
a) $\kappa=\frac{\pi}{4}$;
b) $\kappa=\frac{2\pi}{3}$.

\vspace{1cm}

\noindent
FIG. 8.
Density of states
$\overline{\rho(E)}$
for model (1) with
$J_0=-1,$
$\Omega_0=0.5,$
$\Gamma=0.1.$
a) correlated disorder (3) with $\Gamma=0.1,$ $a=-1.01$;
b) correlated disorder (3) with $\Gamma=0.1,$ $a=1.01$;
c) independent exchange couplings (2) with $\Gamma=0.1$
and transverse fields (4) with $\vert a\vert\Gamma=0.101$;
the density of states for the non-random case $\Gamma=0$
is depicted by dashed curves.

\clearpage

\begin{figure}
\begin{center}
\epsfxsize=100mm
\epsfbox{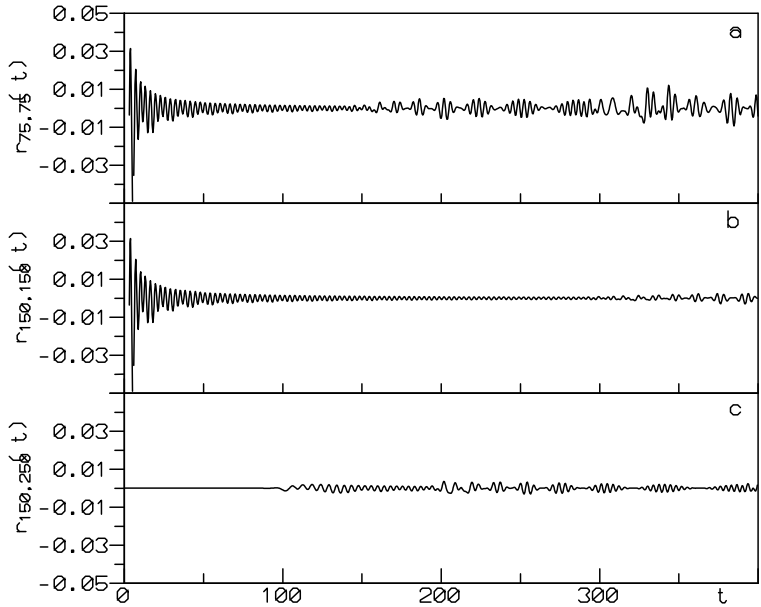}
\end{center}
\vspace{15mm}
\caption[]{
Time dependence of the transverse spin correlation functions
$r_{j,j+n}(t)={\mbox{Re}}
\left[\langle s_{j}^z(t)s_{j+n}^z\rangle
\right.
$
$
\left.
-\langle s_{j}^z\rangle\langle s_{j+n}^z\rangle\right]$
for $J_0=-1$, $\Omega_0=0.5$, $\Gamma=0$ at $\beta=1000$.
a) $N=150$, $j=j+n=75$;
b) $N=300$, $j=j+n=150$;
c) $N=300$, $j=150$, $j+n=250$.
In this figure and in Figs. 2, 4 we plotted only the real parts of
correlation functions since their imaginary parts exhibit
qualitatively the same behaviour.}
\end{figure}

\clearpage

\begin{figure}
\begin{center}
\epsfxsize=100mm
\epsfbox{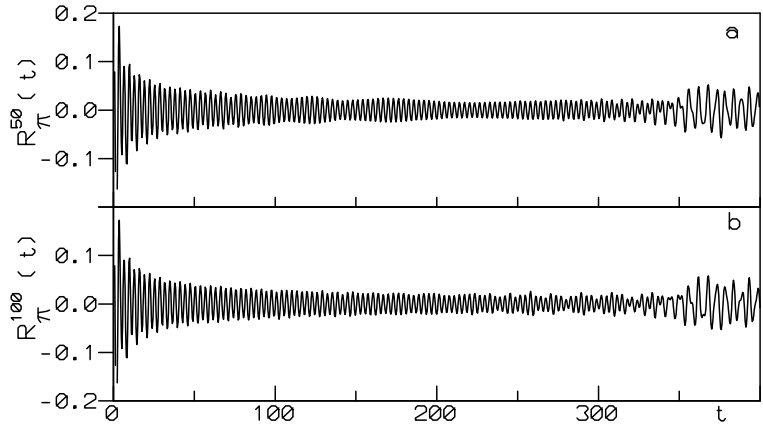}
\end{center}
\vspace{15mm}
\caption[]{
Time dependence of
$R_{\kappa}^{n^{\star}}(t)={\mbox{Re}}
\sum_{n=0,\pm1,\ldots,\pm n^{\star}}
{\mbox{e}}^{{\mbox{i}}\kappa n}
\left[
\langle s_{150}^z(t)s_{150+n}^z\rangle
-\langle s_{150}^z\rangle\langle s_{150+n}^z\rangle
\right],$
$\kappa=\pi$
for $J_0=-1$, $\Omega_0=0.5$, $\Gamma=0$, $N=300$ at $\beta=1000$
for $n^{\star}=50$ (a)
and $n^{\star}=100$ (b).}
\end{figure}

\clearpage

\begin{figure}
\begin{center}
\epsfxsize=100mm
\epsfbox{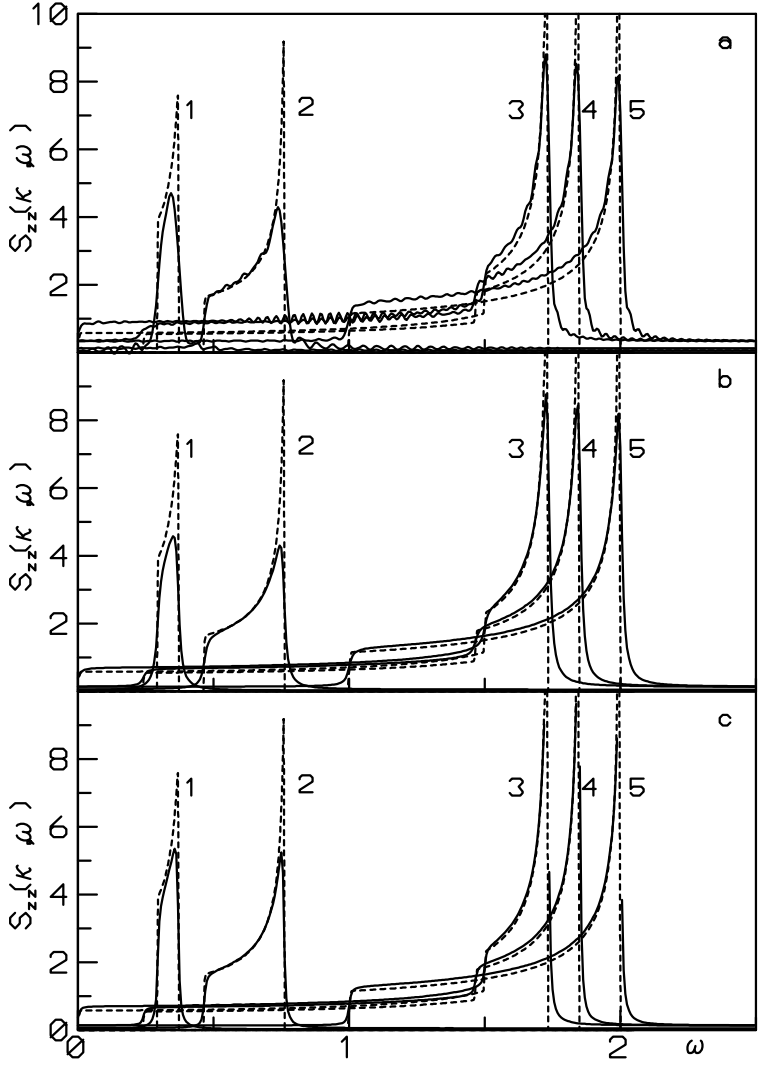}
\end{center}
\vspace{15mm}
\caption[]{
$S_{zz}(\kappa,\omega)$
at $\kappa=\frac{\pi}{8}$ (1),
$\kappa=\frac{\pi}{4}$ (2),
$\kappa=\frac{2\pi}{3}$ (3),
$\kappa=\frac{3\pi}{4}$ (4),
$\kappa=\pi$ (5)
for the uniform chain
with $J_0=-1$, $\Omega_0=0.5$ at $\beta=1000$:
exact results (11), (12) (dashed curves)
versus numerical ones
(solid curves).
a)
$N=280$,
$j=140$,
$n^{\star}=100$,
$t_c=200$, $\epsilon=0.01$;
b)
$N=1800$,
$j=900$,
$n^{\star}=800$,
$t_c=1200$, $\epsilon=0.01$;
c)
$N=1800$,
$j=900$,
$n^{\star}=800$,
$t_c=1200$, $\epsilon=0.005$.}
\end{figure}

\clearpage

\begin{figure}
\begin{center}
\epsfxsize=100mm
\epsfbox{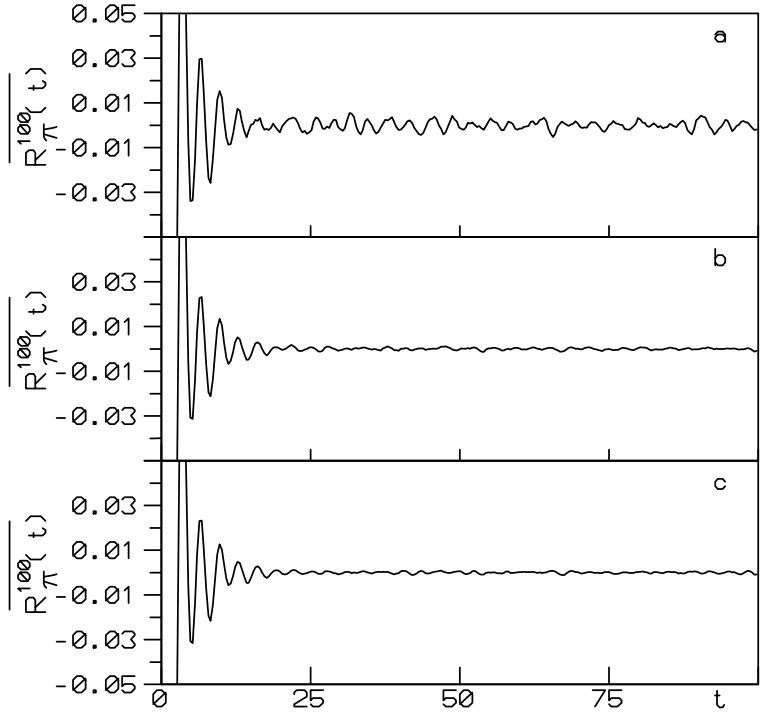}
\end{center}
\vspace{15mm}
\caption[]{
Time dependence of
$\overline{R_{\kappa}^{n^{\star}}(t)},$
$\kappa=\pi$
for $J_0=-1$, $\Omega_0=0.5$,
$\Gamma=0.1$, $a=1.01$,
$N=300$, $n^{\star}=100$ at $\beta=1000$;
the random-averaged quantity comes as the result of
averaging over 1000 realizations (a),
11000 realizations (b),
and 21000 realizations (c).}
\end{figure}

\clearpage

\begin{figure}
\begin{center}
\epsfxsize=50mm
\epsfbox{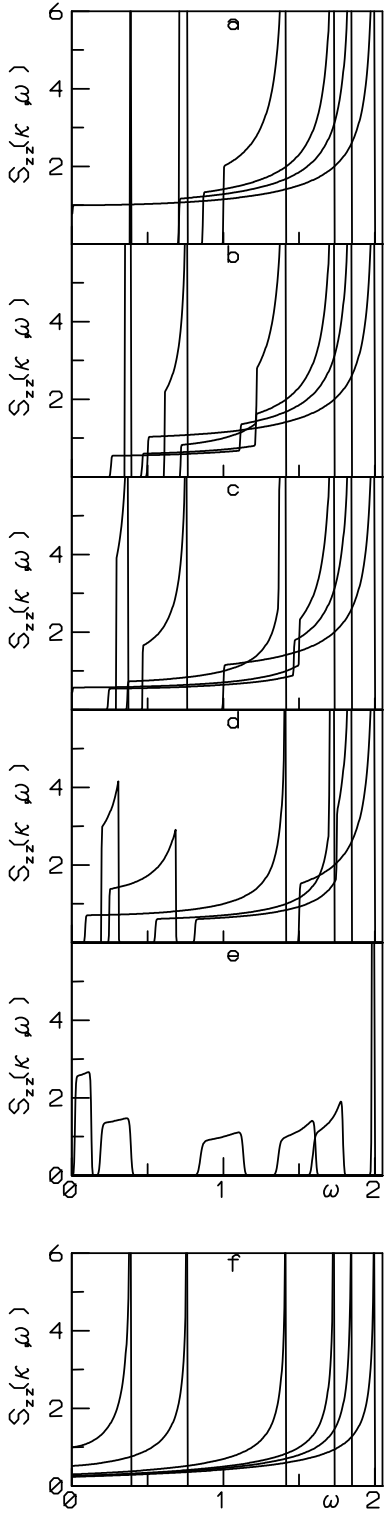}
\end{center}
\vspace{-4mm}
\caption[]{
Frequency dependence of the transverse dynamic structure factor
(11), (12)
for
$J_0=-1$
and
different values of transverse field
$\Omega_0=0$ (a),
$\Omega_0=0.25$ (b),
$\Omega_0=0.5$ (c),
$\Omega_0=0.75$ (d),
$\Omega_0=0.99$ (e)
at different values of wave vector
$\kappa
=\frac{\pi}{8},
\frac{\pi}{4},
\frac{\pi}{2},
\frac{2\pi}{3},
\frac{3\pi}{4},
\pi$
(from left to right)
at $\beta=1000$.
At high temperatures the
frequency profiles of
$S_{zz}(\kappa,\omega)$
are the same for all values of $\Omega_0$ (f).}
\end{figure}

\clearpage

\begin{figure}
\begin{center}
\epsfxsize=50mm
\epsfbox{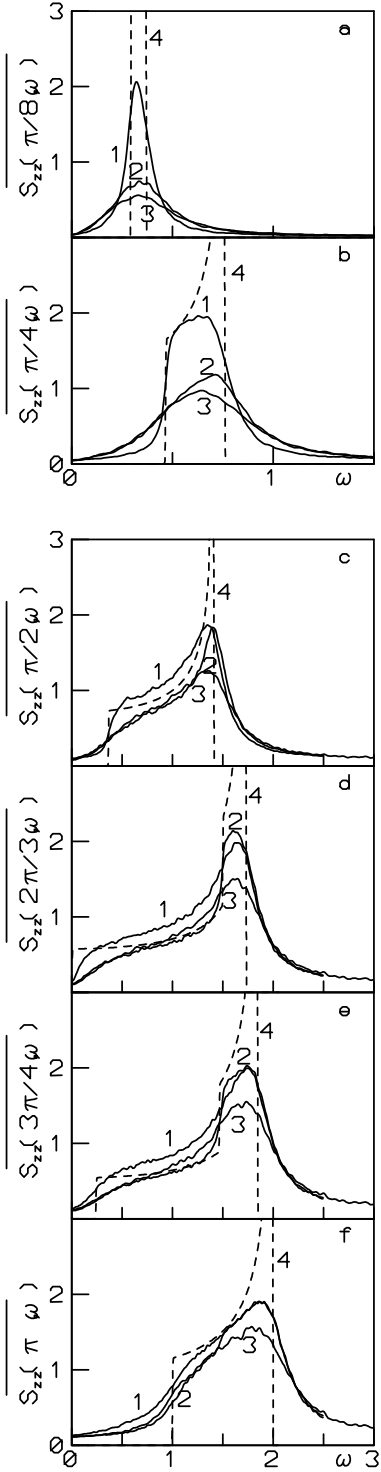}
\end{center}
\vspace{-4mm}
\caption[]{
Frequency dependence of the random-averaged transverse dynamic structure
factor (8) at different values of wave vector
$\kappa=\frac{\pi}{8}$ (a),
$\kappa=\frac{\pi}{4}$ (b),
$\kappa=\frac{\pi}{2}$ (c),
$\kappa=\frac{2\pi}{3}$ (d),
$\kappa=\frac{3\pi}{4}$ (e),
$\kappa=\pi$ (f)
for model (1) with
$J_0=-1,$
$\Omega_0=0.5,$
$\Gamma=0.1$
at $\beta=1000.$
1) correlated disorder (3) with $a=-1.01;$
2) correlated disorder (3) with $a=1.01;$
3) independent exchange couplings and transverse fields,
the latter are distributed according to
probability distribution (4) with
$\vert a\vert\Gamma=0.101;$
4) non-random case $\Gamma=0$ (dashed curves).}
\end{figure}

\clearpage

\begin{figure}
\begin{center}
\epsfxsize=80mm
\epsfbox{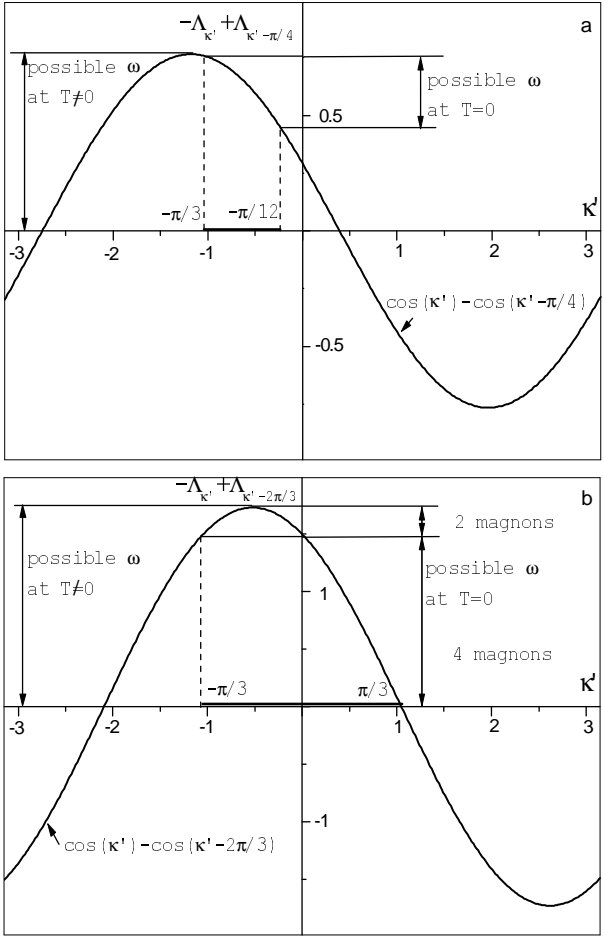}
\end{center}
\vspace{14mm}
\caption[]{
$-\Lambda_{\kappa^{\prime}}
+\Lambda_{\kappa^{\prime}-\kappa}$
versus
$\kappa^{\prime}$
for the non-random chain with $J_0=-1$,
$\Omega_0=0.5$.
a) $\kappa=\frac{\pi}{4}$;
b) $\kappa=\frac{2\pi}{3}$.}
\end{figure}

\clearpage

\begin{figure}
\begin{center}
\epsfxsize=70mm
\epsfbox{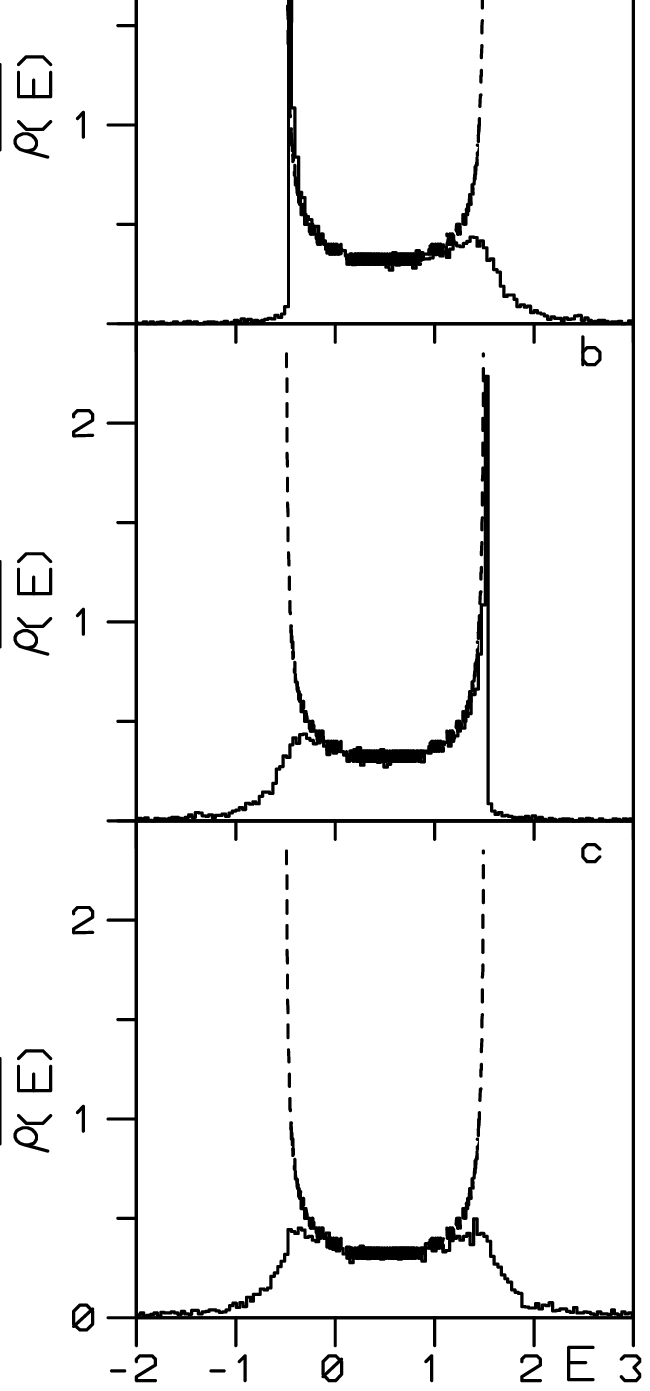}
\end{center}
\vspace{10mm}
\caption[]{
Density of states
$\overline{\rho(E)}$
for model (1) with
$J_0=-1,$
$\Omega_0=0.5,$
$\Gamma=0.1.$
a) correlated disorder (3) with $\Gamma=0.1,$ $a=-1.01$;
b) correlated disorder (3) with $\Gamma=0.1,$ $a=1.01$;
c) independent exchange couplings (2) with $\Gamma=0.1$
and transverse fields (4) with $\vert a\vert\Gamma=0.101$;
the density of states for the non-random case $\Gamma=0$
is depicted by dashed curves.}
\end{figure}


\begin{thebibliography}{99}
\bibitem{fn1}
In our numerical calculations we put
$\Omega_1=\Omega_0+a\left(J_1-J_0\right)$
and
$\Omega_{N}=\Omega_0+a\left(J_{N-1}-J_0\right)$.
\bibitem{fn2}
Formula (9) can be also used
for the numerical calculation of the transverse
dynamic structure factor.
\end{thebibliography}
\end{document}